# Penganggaran Impak Pelaksanaan CBP ke atas Kos Pengeluaran dan Kos Sara Hidup di Malaysia

*(Estimating the Impact of GST Implementation on Cost of Production and Cost of Living in Malaysia)*

**Asan Ali Golam Hassan**
Universiti Teknologi Malaysia

**Mohd Yusof Saari**
**Chakrin Utit**
**Azman Hassan**
Universiti Putra Malaysia

**Mukaramah Harun**
Universiti Utara Malaysia

*ABSTRAK*

*Pelaksanaan Cukai Barangan dan Perkhidmatan (CBP) seringkali dikaitkan sebagai punca utama kepada kenaikan harga barangan dan perkhidmatan. Objektif utama kajian ini ialah menganggar sejauhmana pelaksanaan CBP memberi impak kepada kos pengeluaran yang seterusnya memberi implikasi kepada kos sara hidup isi rumah. Untuk mencapai objektif tersebut, model harga input-output telah dibangunkan dengan pengubahsuaian bagi mengambil kira tiga kadar cukai yang berbeza iaitu pembekalan berkadar standard, pembekalan berkadar sifar dan pembekalan dikecualikan. Dapatan kajian mendapati bahawa pelaksanaan CBP cenderung memberi faedah kepada ekonomi. CBP berpotensi untuk menurunkan harga barangan dan perkhidmatan sebanyak 7.4% berbanding hanya 2.0% peningkatan harga yang dijangka. Kesan penurunan tingkat harga sebanyak 5.4% ini bukan sahaja meningkatkan kuasa beli perbelanjaan isi rumah malah berpotensi untuk merapatkan jurang penggunaan di antara kumpulan etnik dan kelas pendapatan. Walaubagaimanapun, faedah sebenar kepada ekonomi bergantung kepada kadar pematuhan industri kepada sistem percukaian CBP.*

*Kata kunci: Cukai Barangan dan Perkhidmatan (CBP); input-output; kos pengeluaran; kos sara hidup*

*ABSTRACT*

*The implementation of Goods and Services Tax (GST) is often attributed as the main cause of the rising prices of goods and services. The main objective of this study is to estimate the extent of GST implementation impact on the costs of production, which in turn have implication on households' living costs. To achieve these objectives, the input-output price model is developed with modifications to take into account three different tax categories namely standard-rated supplies, zero-rated supplies and exempt supplies. The study finds that the implementation of the GST tends to benefit the economy. GST has the potential to reduce the prices of goods and services by 7.4% compared to only 2.0% expected price increase. The impact of this 5.4% price reduction not only increases the purchasing power of household but also has the potential to narrow the consumption gap between ethnic groups and income classes. However, the actual benefits to the economy depending on the compliance rates of industries to the GST system.*

## PENGENALAN

Pada 1 April 2015, Malaysia telah memperkenalkan sistem percukaian baharu iaitu, Cukai Barangan dan Perkhidmatan (CBP) yang menggantikan sepenuhnya sistem Cukai Jualan dan Perkhidmatan (CJP). CBP merupakan cukai kepenggunaan yang berasaskan kepada konsep nilai ditambah. Di bawah sistem CBP, cukai dikenakan ke atas barangan dan perkhidmatan di setiap peringkat pengeluaran dan pengedaran dalam rantaian pembekalan. Sebagai permulaan, kadar CBP adalah 6% berbanding dengan kadar cukai jualan terdahulu yang dilaksanakan pada kadar 5% hingga 10% dan cukai perkhidmatan pada kadar 6%. Pada dasarnya, pelaksanaan CBP merupakan satu daripada pendekatan jangka panjang kerajaan bagi meningkatkan hasil pendapatan negara. Cukai ini lebih komprehensif kerana ia mencakupi hampir kesemua barangan dan perkhidmatan yang terdapat di pasaran berbanding cukai jualan dan perkhidmatan yang tertakluk kepada paras nilai yang tertentu sahaja. Di dalam komuniti ASEAN, Malaysia merupakan negara yang kelapan yang telah menguatkuasakan pelaksanaan CBP. Indonesia telah melaksanakan cukai penggunaan tidak langsung ini semenjak tahun 1984, diikuti oleh Thailand (1992), Singapura (1993), Filipina (1998), Kemboja dan Vietnam (1999) dan Laos (2009).

Pelaksanaan sistem percukaian CBP mampu untuk meningkatkan kecekapan ekonomi yang dapat dilihat daripada beberapa sudut. Pertama, CBP cenderung untuk mengurangkan kos perniagaan kepada pihak pengeluar. Di bawah CJP, peniaga tidak dapat menuntut sepenuhnya cukai yang dibayar ke atas input pengeluaran. Di bawah sistem CBP kos pengeluaran dijangka lebih rendah disebabkan pengeluar boleh mendapatkan semula cukai input ke atas bahan mentah di setiap peringkat rantaian pembekalan. Kedua, peningkatan penyertaan perniagaan kepada sistem CBP membantu meningkatkan pematuhan ke atas pembayaran cukai oleh sektor-sektor ekonomi tidak formal yang terdiri daripada aktiviti-aktiviti perniagaan yang tidak berdaftar dengan pihak berwajib. Hal ini kerana untuk menuntut cukai yang dibayar ke atas input pengeluaran, sektor-sektor ekonomi tidak formal perlu mendaftar perniagaan mereka terlebih dahulu dengan pihak berwajib. Ketiga, CBP adalah sistem percukaian yang lebih adil kerana ia dikenakan secara sama rata ke atas kesemua pengeluar yang terlibat di dalam sesebuah rantaian pembekalan.

Walaupun pelaksanaan CBP mampu meningkatkan kecekapan ekonomi, kajian literatur jelas menunjukkan bahawa CBP juga berpotensi meningkatkan harga barangan dan perkhidmatan yang seterusnya memberi impak kepada kos sara hidup isi rumah. Contohnya, di Kanada, Dungan dan Wilson (1989) telah menunjukkan CBP meningkatkan harga di antara 1.5% sehingga 2.0%. Di Hungary, Gabriel dan Reiff (2010) telah mendapati peningkatan sebanyak 3.0% kadar CBP membawa kepada inflasi sebanyak 2.1%. Di Singapura, pelaksanaan CBP telah meningkatkan kadar inflasi kepada 3.6% dan kadar ini adalah jauh lebih rendah daripada unjuran yang dilakukan oleh pihak kerajaan Singapura yang menganggarkan kadar inflasi akan meningkat kepada 5.5% (Jenkins dan Khadka 1998). Manakala di Thailand pula, Ruangmalai (1993) mendapati CBP berpotensi untuk meningkatkan kadar harga dan peningkatan tertinggi yang dicatatkan adalah sebanyak 5.0% iaitu untuk produk farmaseutikal. Justeru itu, amat penting sekali kajian khusus penilaian impak CBP ke atas harga barangan dan perkhidmatan dan kos hidup isi rumah di Malaysia dilakukan.

Objektif utama kajian ini ialah menganggar sejauhmana pelaksanaan CBP di Malaysia memberi kesan kepada kos pengeluaran dan kos sara hidup isi rumah. Untuk mencapai objektif tersebut, model harga input-output telah dibangunkan dengan pengubahsuaian bagi mengambil kira tiga kadar cukai yang berbeza iaitu pembekalan berkadar standard, pembekalan berkadar sifar dan pembekalan dikecualikan. Perlu ditegaskan di sini bahawa skop kajian ini adalah terhad kepada penilaian impak CBP sahaja. Ini bermakna kajian tidak mengambil kira pengaruh dasar-dasar tekanan harga lain seperti upah minimum, rasionalisasi subsidi petroleum dan kekusutan nilai matawang Ringgit yang boleh meningkatkan harga barangan dan perkhidmatan. Permodelan dasar-dasar tekanan harga lain bersama-sama dengan pembolehubah CBP memerlukan kepada pembangunan model yang komprehensif dan memerlukan data yang besar. Lebih menyukarkan lagi apabila pengaruh ketiga-tiga dasar di atas sukar untuk dikawal dan dipengaruhi oleh faktor-faktor luaran.

Terdapat tiga sumbangan utama kajian ini kepada literatur ekonomi Malaysia. Pertama, penilaian impak CBP ke atas harga barangan dan perkhidmatan, dan kos hidup isi rumah adalah yang pertama dijalankan dengan menggunakan data yang terkini. Jadual input-output dan Tinjauan Perbelanjaan Isi rumah (HES) yang terkini telah digunakan untuk menjalankan analisis. Kedua, kajian ini menyediakan analisis impak harga dan kos hidup secara terperinci. Di mana impak ke atas harga diperinci sehingga 124 jenis barangan dan perkhidmatan, dan impak kos sara hidup dinilai mengikut kumpulan etnik dan kelas pendapatan. Ketiga, model harga input-output telah dibangunkan secara spesifik untuk mengambil kira tiga kadar cukai yang berbeza iaitu pembekalan berkadar standard, pembekalan berkadar sifar dan pembekalan dikecualikan.

Kertas ini disediakan mengikut lima bahagian. Bahagian 2 merumus dapatan kajian terdahulu daripada pelbagai negara berkaitan dengan impak CBP ke atas harga dan kos sara hidup. Bahagian 3 memperinci metodologi dan data yang digunakan. Bahagian 4 membentang dapatan utama kajian yang melihat impak CBP ke atas tingkat harga barangan dan perkhidmatan dan kos sara hidup isi rumah. Akhirnya, Bahagian 5 merumus dapatan kajian.

## KAJIAN LITERATUR

Rumusan daripada kajian terdahulu kebanyakannya menunjukkan bahawa pelaksanaan CBP berpotensi untuk meningkatkan harga barangan dan perkhidmatan dan seterusnya memberi implikasi kepada kos sara hidup isi rumah. Saiz impak yang akan dibawanya adalah berbeza kerana ia bergantung kepada struktur ekonomi negara yang berkenaan dan turut bergantung kepada faktor jangka masa (harga cenderung untuk stabil dalam jangka masa panjang).

Di Amerika Syarikat, Carroll et al. (2010) mengunjurkan bahawa pelaksanaan CBP akan menyebabkan *Federal Reserve Board* untuk meningkatkan jumlah wang di pasaran dan sekaligus membawa kepada masalah inflasi. Namun begitu,

*Penganggaran Impak Pelaksanaan CBP ke atas Kos Pengeluaran dan Kos Sara Hidup di Malaysia* 17perkara ini cuma akan berlaku pada peringkat awal pelaksanaan CBP sahaja dan ia tidak mencerminkan peningkatan berterusan dalam kadar inflasi pada tahun-tahun yang berikutnya. Harus diingatkan juga bahawa kajian ini adalah bertujuan untuk menganggarkan impak perlaksanaan CBP sahaja dan sehingga kini Amerika Syarikat masih belum melaksanakan sistem percukaian CBP. Di Kanada, Dungan dan Wilson (1989) menunjukkan CBP meningkatkan harga di antara 1.5% sehingga 2%. Huang dan Liu (2012) pula telah mendapati kos pinjaman rumah mampu milik di Australia telah meningkat dengan pelaksanaan CBP.

Dalam kajian pengaruh CBP ke atas kestabilan harga di Nigeria, Ikpe dan Nteegah (2013) mendapati bahawa CBP membawa tekanan kepada tingkat harga akibat daripada tekanan kos input perantaraan. Kajian oleh Jenkins dan Kuo (2000) di Nepal turut mendapati berlakunya kenaikan kepada harga barangan dan perkhidmatan kepada pengguna akhir sekiranya cukai dikenakan ke atas barangan perantaraan. Hal ini terbukti apabila perbelanjaan isi rumah di kawasan luar bandar didapati meningkat daripada 7% (1985) kepada 12% (1994). Perkara ini turut disokong oleh Bhupalan (2011) yang menyatakan pelaksanaan CBP akan menyebabkan kenaikan harga barangan dan perkhidmatan yang akan memberi kesan kepada isi rumah dan kos pengeluaran.

Pelaksanaan CBP turut dilihat menyumbang kepada masalah inflasi di Hungary (Gabriel & Reiff 2010). Dapatan kajian menunjukkan inflasi telah meningkat sebanyak 6.7% daripada 2.3% pada April 2006 kepada 9% pada Mac 2007. Namun begitu, kesan pelaksanaan CBP akan dapat diminimakan sekiranya dikuatkuasakan dalam masa yang sesuai. Berdasarkan kajian yang pernah dijalankan ke atas Britain dan Jerman, ahli ekonomi beranggapan bahawa pelaksanaan CBP akan memberikan kesan yang paling minima ke atas kadar inflasi negara sekiranya ia dilaksanakan ketika tempoh keadaan ekonomi yang bergerak perlahan (Palil & Ibrahim 2011).

Untuk melihat dengan lebih jelas kesan pelaksanaan CBP di negara-negara terpilih dalam tempoh lima tahun pertama pelaksanaannya, data tingkat harga bagi negara-negara terpilih telah dikutip dan ditunjukkan pada Jadual 1. Perlu ditegaskan di sini bahawa maklumat pada Jadual 1 hanya menunjukkan tren pergerakan tingkat harga. Walaupun CBP mungkin mempengaruhi sebahagian besar tingkat harga, faktor-faktor lain yang memberi tekanan harga juga mempengaruhi pergerakan tingkat harga tersebut.

Jika dilihat kepada kes negara maju, Singapura yang mula melaksanakan sistem CBP pada tahun 1993 pada kadar permulaan 3.0%, peratusan kadar inflasi secara purata daripada tahun 1993 ke 1997 adalah di tahap 2.1% sahaja. Ini menunjukkan bahawa

JADUAL 1. Kadar inflasi dalam tempoh lima tahun pelaksanaan CBP di negara-negara terpilih

| | Kadar permulaan CBP (%) | Kadar inflasi (%) | | | | |
|---|---|---|---|---|---|---|
| | | Tahun 1 | Tahun 2 | Tahun 3 | Tahun 4 | Tahun 5 |
| *Negara berpendapatan rendah* | | | | | | |
| Madagascar (1994) | 20.0 | 38.94 | 49.08 | 19.76 | 4.49 | 6.21 |
| Nepal (1997) | 13.0 | 4.01 | 11.24 | 7.45 | 2.48 | 2.69 |
| Cameroon (1999) | 19.0 | 1.88 | 1.23 | 4.42 | 2.83 | 0.62 |
| Nigeria (1993) | 5.0 | 57.17 | 57.03 | 72.84 | 29.27 | 8.53 |
| Bangladesh (1991) | 15.0 | 6.38 | 3.63 | 3.01 | 5.31 | 10.30 |
| India (2005) | 12.5 | 4.25 | 5.79 | 6.39 | 8.32 | 10.83 |
| *Negara berpendapatan sederhana* | | | | | | |
| Serbia (2004) | 20.0 | 11.03 | 16.12 | 11.72 | 6.40 | 12.41 |
| Filipina (1998) | 10.0 | 9.27 | 5.95 | 3.95 | 5.35 | 2.72 |
| Thailand (1992) | 7.0 | 4.14 | 3.31 | 5.04 | 5.82 | 5.81 |
| Hungary (1988) | 27.0 | 15.75 | 9.99 | 9.80 | 9.15 | 5.27 |
| China (1994) | 17.0 | 24.17 | 17.07 | 8.33 | 2.81 | -0.77 |
| Indonesia (1984) | 10.0 | 10.49 | 4.74 | 5.82 | 9.29 | 8.07 |
| *Negara berpendapatan tinggi* | | | | | | |
| Singapura (1993) | 3.0 | 2.29 | 3.10 | 1.72 | 1.38 | 2.00 |
| Kanada (1991) | 5.0 | 5.64 | 1.49 | 1.87 | 0.17 | 2.15 |
| Jepun (1989) | 5.0 | 2.27 | 3.04 | 3.30 | 1.71 | 1.27 |
| Australia (2000) | 10.0 | 2.34 | 2.65 | 1.81 | 1.36 | 2.06 |

*Nota*: ( ) merujuk kepada tahun pelaksanaan CBP di negara-negara berkenaan
*Sumber*: Jabatan Kastam Diraja Malaysia (2014) dan World Bank (2015)

tahap inflasi masih berada di paras normal dan tidak akan memberikan kesan yang besar kepada prestasi keseluruhan ekonomi negara tersebut. Bagi kes negara berpendapatan sederhana seperti Thailand yang mempunyai struktur ekonomi yang hampir serupa dengan Malaysia, kadar inflasi jatuh pada tahun kedua pelaksanaannya dan kemudian meningkat dengan agak ketara pada tiga tahun berikut. Manakala bagi Filipina dan Indonesia, kenaikan harga yang ketara berlaku pada tahun pertama pelaksanaannya dan kemudian jatuh pada tahun-tahun berikutnya.

Peningkatan dalam harga barangan dan perkhidmatan akibat CBP akan meningkatkan kos perbelanjaan isi rumah dan seterusnya memberi implikasi kepada kos sara hidup. Kos sara hidup akan lebih terjejas jika pergerakan pendapatan isi rumah tidak setara dengan peningkatan kos perbelanjaan. Berdasarkan kajian oleh Caspersen dan Metcalf (1993), CBP adalah cukai penggunaan yang boleh menyebabkan golongan miskin mempunyai nisbah penggunaan pendapatan yang lebih tinggi ke atas barang dan perkhidmatan berbanding golongan kaya. Selain itu, pembaharuan cukai di Serbia daripada cukai pendapatan pengguna kepada cukai penggunaan individu pula menyebabkan jumlah penggunaan individu menurun sebanyak 3.4% manakala jumlah kemiskinan meningkat daripada 6.9% kepada 7.4% dalam jangka masa pendek (Matkovic & Mijatovic 2011). Piggott dan Whalley (2001) pula menunjukkan bahawa peluasan asas cukai penggunaan boleh mengurangkan kebajikan masyarakat. Ini disokong oleh Hooper dan Smith (1997) yang menyatakan CBP akan memberi kesan yang mendalam kepada golongan berpendapatan rendah berbanding dengan golongan yang berpendapatan lebih tinggi.

Sebahagian daripada kajian terdahulu mendapati CBP juga membawa kesan positif kepada ekonomi. Contohnya, Vasanthagopal (2011) mendapati pelaksanaan CBP di India telah memihak kepada rakyat yang berpendapatan rendah. Hal ini berlaku kerana bahan makanan asas yang dahulunya tertakluk di bawah cukai jualan telah dikecualikan daripada CBP. Kajian tersebut juga menunjukkan CBP akan meningkatkan harga hasil pertanian di antara 0.61% hingga 1.18%. Perkara ini akan menguntungkan para petani atau golongan berpendapatan rendah di India. Gupta (2014) turut mendapati kadar inflasi akan menurun selepas satu jangka masa kerana kadar CBP yang dikenakan adalah seragam. Perkara ini akan menguntungkan para peniaga dan juga pengguna disebabkan bebanan kos pengeluaran dan sara hidup yang berkurangan.

## METODOLOGI DAN DATA

Kajian ini menggunakan teknik input-output sebagai metodologi utama untuk mencapai objektif kajian. Input-output merupakan teknik penilaian impak yang selalu digunakan oleh penyelidik dan penganalisis dasar ekonomi di seluruh dunia. Teknik ini popular untuk penilaian impak kerana kemampuannya untuk mengambil kira kebergantungan ke atas rantaian bekalan bagi sektor-sektor di dalam ekonomi. Rajah 1 meringkaskan kerangka methodologi yang digunakan di dalam kajian ini.

Berdasarkan Rajah 1, metodologi kajian boleh diringkaskan kepada tiga langkah utama. Langkah pertama kajian ialah mengenalpasti pemboleh ubah CBP di dalam jadual input-output. Di dalam jadual input-output tahun asas 2010, pembolehubah CBP terdapat pada item "*taxes on products (domestic)*". *Taxes on products (domestic)* merujuk kepada cukai jualan dan perkhidmatan yang diamalkan pada masa kini. Penggantian daripada cukai jualan dan perkhidmatan kepada CBP bermakna pembolehubah *taxes on products*

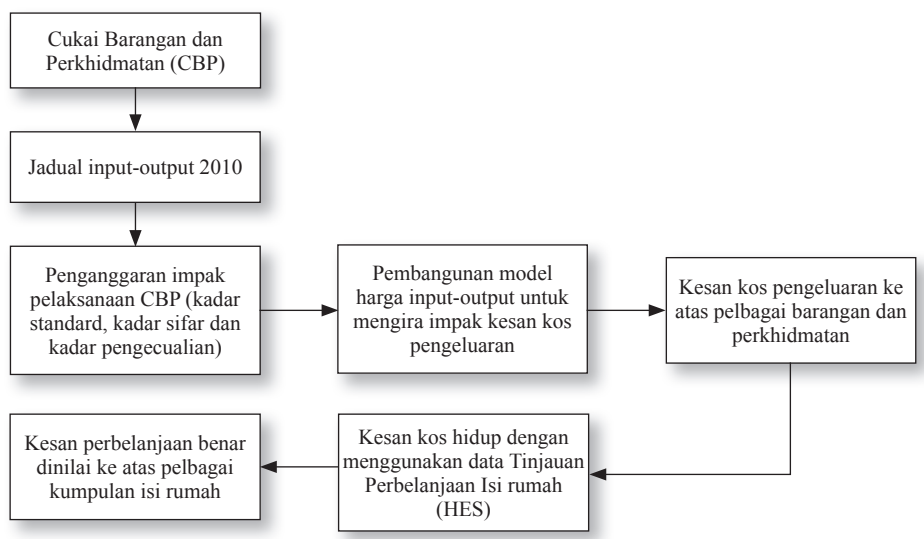

RAJAH 1. Kerangka metodologi penganggaran impak pelaksanaan CBP

*(domestic)* akan menjadi pembolehubah sasaran di dalam analisis kajian.

Setelah mengenalpasti pembolehubah sasaran, langkah kedua ialah membina model harga input-output untuk menganggar impak cukai CBP. Terdapat tiga kadar CBP yang diperkenalkan iaitu pembekalan berkadar standard, pembekalan berkadar sifar dan pembekalan dikecualikan. Ketiga-tiga kadar CBP ini akan memberi impak yang berbeza kepada kos pengeluaran. Senarai barangan dan perkhidmatan yang terkandung di bawah kategori pembekalan berkadar sifar dan pembekalan dikecualikan diberikan oleh pihak Jabatan Kastam Diraja Malaysia. Oleh hal yang demikian, permodelan input-output juga perlu diubahsuai dengan mengambil kira perbezaan mekanisma ketiga-tiga kadar CBP tersebut.

Langkah ketiga ialah mengira sejauhmana kesan perubahan harga barangan dan perkhidmatan akan memberi implikasi ke atas kos sara hidup isi rumah. Untuk mendapatkan kesan kos sara hidup isi rumah, hasil dapatan daripada model harga input-output akan dihubungkait dengan data daripada Tinjauan Perbelanjaan Isi rumah (HES). Impak ke atas isi rumah ditumpukan kepada kumpulan isi rumah mengikut kumpulan etnik (Bumiputera, Cina, India dan lain-lain) dan kelas pendapatan. Penumpuan analisis kepada kumpulan etnik dan kelas pendapatan dilakukan kerana kedua-dua pembahagian kumpulan ini menjadi sasaran utama pihak kerajaan di dalam pelan pembangunan ekonomi.

Perbincangan seterusnya memperinci metodologi penganggaran kesan perubahan kos pengeluaran industri dan kesan perubahan kos sara hidup (yang diukur melalui perubahan perbelanjaan). Bahagian 3.1 memperinci model harga input-output yang digunakan untuk menganggar kesan perubahan kos pengeluaran. Bahagian 3.2 memperinci permodelan kesan perubahan kos pengeluaran ke atas kos sara hidup. Bahagian 3.3 pula memperinci data yang digunakan untuk permodelan pada Bahagian 3.1 dan 3.2.

## MODEL HARGA INPUT-OUTPUT

Model harga input-output digunakan untuk menganggar kesan perubahan cukai tidak langsung ke atas perubahan kos pengeluaran setiap sektor ekonomi. Jadual input-output yang terkini iaitu pada tahun asas 2010 digunakan untuk membuat analisis. Jadual ini dikeluarkan oleh Jabatan Perangkaan Malaysia pada tahun 2014 yang meliputi 124 sektor (Jabatan Perangkaan Malaysia 2014). Setiap sektor di dalam jadual input-output mewakili sekumpulan aktiviti ekonomi yang dikelaskan mengikut *Malaysia Standard Industrial Classification* (MSIC 2008).

Model input-output menganalisis kebergantungan di antara pelbagai sektor ekonomi di dalam penggunaan dan penjualan output. Output yang dikeluarkan oleh sesebuah sektor digunakan oleh sektor-sektor lain sebagai input kepada proses pengeluaran mereka. Justeru itu, perubahan kos pengeluaran sesebuah sektor akibat daripada faktor-faktor tertentu (seperti perubahan struktur cukai) akan turut menjejaskan kos pengeluaran sektor-sektor ekonomi lain. Kebergantungan di antara pelbagai sektor ekonomi boleh diwakili oleh persamaan berikut:

$$\mathbf{x} = \Sigma\mathbf{Z} + \mathbf{f} + \mathbf{e} \qquad (1)$$

di mana, $\mathbf{x}$ ialah vektor-lajur output. Sebahagian output yang dikeluarkan oleh sesebuah sektor akan digunakan sebagai input perantaraan domestik, diwakili oleh matrik $\mathbf{Z}$ (iaitu input yang diminta oleh sektor $j$ sebagai input perantaraan daripada sektor $i$), sebahagiannya lagi digunakan oleh sektor permintaan akhir, diwakili oleh vektor-lajur $\mathbf{f}$ (seperti isi rumah dan kerajaan) dan selebihnya untuk tujuan eksport, diwakili oleh vektor-lajur $\mathbf{e}$. Dalam permodelan standard, persamaan di atas boleh ditransformasi dan diselesaikan seperti berikut:

$$\begin{aligned} x &= Ax + (f+e) \\ &= (I-A)^{-1}(f+e) = L(f+e) \end{aligned} \qquad (2)$$

di mana, I adalah matrik identiti, matrik $A (A = Z\hat{x}^{-1})$ merupakan pekali input-output dan L adalah matrik songsang Leontief. Bagi setiap sektor $j$, pekali matrik songsang Leontief mewakili keperluan output langsung (*direct*) dan tak langsung (*indirect*) untuk memenuhi setiap unit permintaan akhir. Model (2) juga dikenali sebagai model-kuantiti yang mana di dalam model ini, kuantiti input yang digunakan di dalam pengeluaran menjadi pembolehubah manakala harga diandaikan tetap. Untuk memastikan andaian harga tetap adalah realistik, model ini mengandaikan juga terdapat lebihan penawaran (seperti lebihan penawaran buruh) dan wujudnya hubungan linear di antara pembolehubah.

Dualiti kepada model-kuantiti ialah model-harga atau model-tolakan kos. Model-harga sangat berguna untuk analisis harga dan kos seperti cukai, duti import dan upah buruh kerana model ini mengandaikan kuantiti input yang digunakan adalah tetap manakala harga atau kos dianggap sebagai pembolehubah. Model-harga input-output boleh diwakili seperti persamaan berikut:

$$\begin{aligned} p &= A'p + 1 + v + m + t \\ &= (I - A')^{-1}(1 + v + m + t) = L'(1 + v + m + t) \end{aligned} \qquad (3)$$

di mana, $p$ ialah vektor-lajur harga *normalized* (iaitu harga bersamaan dengan nilai 1 untuk *baseline*); $A'$ adalah transposisi matrik pekali input-output; dan 1, $v$, $t$ dan masing-masing mewakili vektor-lajur pekali pendapatan buruh (pendapatan buruh per unit output), pekali pendapatan kapital (pendapatan kapital per unit output), pekali import (import per unit output) dan pekali cukai tidak langsung (cukai tidak langsung per unit output). Di dalam model (3) 1, $v$, $t$ dan adalah pembolehubah eksogenous. Pada *baseline*, harga bagi setiap pembolehubah eksogenous ditetapkan pada nilai 1, maka model (3) boleh diringkas seperti berikut:

$$p = L'(1 + v + m + t) \quad (4)$$

Di dalam kajian ini, pembolehubah-pemboleubah 1, $v$, dan $m$ dianggap tetap dan hanya $t$ dianggap sebagai pembolehubah (iaitu hanya $t$ yang menentukan perubahan $p$). Di dalam model ini, bila mana tiada perubahan ke atas kadar cukai tidak langsung, ia bermakna nilai pekali cukai tidak langsung bersamaan dengan $t$. Jika berlaku perubahan kadar cukai tidak langsung, perubahan ini diterjemahkan seperti berikut: $\Delta t = \Delta t \otimes pt$. Di mana $\otimes$ merujuk kepada *Hadamard product*, iaitu pendaraban *cell-by-cell* dan $\Delta p_t$ merujuk kepada perubahan kadar cukai. Pada *baseline*, $p_t$ bersamaan dengan 1 dan andaikan cukai meningkat sebanyak 10%, maka nilai $\Delta p_t$ adalah 1.10. Dengan menggunakan konsep *price pass-through*, kenaikan kos pengeluaran akibat peningkatan kadar cukai akan diterjemahkan kepada kenaikan harga barangan dan perkhidmatan. Justeru itu, kesan peningkatan cukai tidak langsung kepada harga barangan dan perkhidmatan semua sektor ekonomi dapat diwakili seperti persamaan berikut:

$$\Delta p = L'(1 + v + m + \Delta t) \quad (5)$$

Untuk kajian ini, model (5) tidak boleh digunakan secara terus dan memerlukan kepada pengubahsuaian. Dua faktor menyumbang kepada pengubahsuaian ini. Pertama, model (5) hanya boleh digunakan jika sistem percukaian CBP di Malaysia berkadar standard untuk semua barangan dan perkhidmatan. Sebaliknya terdapat tiga kadar percukaian CBP di Malaysia, iaitu pembekalan berkadar standard, pembekalan berkadar sifar dan pembekalan dikecualikan. Ketiga-tiga kadar CBP ini akan memberi impak yang berbeza kepada kos pengeluaran. Oleh yang demikian, permodelan harga input-output juga perlu diubahsuai dengan mengambil kira perbezaan mekanisma ketiga-tiga kadar CBP tersebut. Untuk mengambil kira ketiga-tiga kadar CBP ini, matrik songsang Leontief perlu diubahsuai seperti berikut.

$$\Delta p = (I - A'\hat{B})^{-1}(1 + v + m + \Delta t) = L'(1 + v + m + \Delta t) \quad (6)$$

di mana, $\hat{B}$ mewakili matrik diagonal yang digunakan untuk mengubahsuai pekali input-output, $A'$, mengikut kategori CBP. Nilai *off-diagonal* matrik $\hat{B}$ bersamaan dengan 1 jika pengeluaran sektor berkenaan berkadaran standard dan 0 jika ianya berkadaran sifar. Kami memperincikan pengiraan model (6) dengan mengunakan contoh hipotetikal matrik 3 × 3 pada Lampiran 1.

Kedua, pekali cukai tidak langsung merujuk kepada cukai bagi setiap unit output yang dikeluarkan (secara spesifiknya ia merujuk kepada cukai jualan dan perkhidmatan, CJP). Sedangkan CBP merupakan cukai yang bersandarkan kepada nilai ditambah. Ini bermaksud bahawa simulasi 6% CBP tidak boleh terarah kepada pekali cukai tidak langsung, $t$, dan nilai pekali cukai tidak langsung ini memerlukan juga kepada pengubahsuaian. Pengubahsuaian ini melibatkan dua langkah mudah. Pertama, kami mengira nilai pekali cukai tidak langsung kepada nilai ditambah pada *baseline* (sebelum simulasi CBP). Kami wakilkan nilai pekali ini dengan simbol $u$. Seterusnya, nilai pekali cukai tidak langsung, $u$, ini digantikan dengan dengan kadar 6%. Maka, model lengkap impak CBP boleh dirumuskan seperti persamaan berikut:

$$\Delta p = (I - A'\hat{B})^{-1}(1 + v + m + u) = L'(1 + v + m + u) \quad (7)$$

Dua perkara utama berkaitan dengan model input-output perlu diperjelaskan. Pertama, model input-output adalah linear, di mana struktur penggunaan input oleh setiap sektor adalah tetap. Implikasi daripada andaian ini adalah kenaikan kos pengeluaran akan dilepaskan sepenuhnya kepada pengguna akhir (*price pass-through*). Dalam analisis ke atas CBP, andaian ini adalah sah kerana pengeluar akan mengenakan kadar cukai kepada pengguna akhir. Kedua, jadual input-output yang digunakan adalah untuk tahun rujukan 2010. Wujud persoalan sejauhmanakah struktur pengeluaran yang diwakili oleh pekali input-output adalah stabil. Justeru itu, analisis sensitiviti ke atas pekali input perlu dilakukan. Untuk tujuan itu, jadual input-output tahun 2005 dan 2010 digunakan untuk melihat perbezaan pekali input-output (data 2005 dikira dalam harga malar 2010). *Mean absolute deviation* (MAD, rujuk Miller & Blair 2009) dikira dan dapatan menunjukkan MAD adalah 0.023 (nilai MAD yang mendekati sifar menunjukkan kestabilan yang tinggi). Jadi, penggunaan jadual input-output tahun rujukan 2010 adalah realistik.

## MODEL PENGGUNAAN ISI RUMAH

Setelah perubahan harga barangan dan perkhidmatan ke atas 124 sektor ekonomi dikira pada model (7), langkah seterusnya ialah menilai sejauhmana perubahan harga barangan dan perkhidmatan mempengaruhi perbelanjaan isi rumah. Untuk tujuan ini, perubahan perbelanjaan setiap kumpulan isi rumah bagi setiap barangan dan perkhidmatan perlu dikira. Jika berlaku peningkatan ke atas harga barangan dan perkhidmatan, kita perlu mengira perbelanjaan tambahan yang diperlukan oleh isi rumah untuk mengekalkan kuasa beli sebelum tempoh perubahan cukai. Perubahan perbelanjaan boleh dikira seperti berikut:

$$\Delta E = \hat{E} - E = \hat{P}Q - PQ \quad (8)$$

di mana $E$ and $\hat{E}$ adalah matrik perbelanjaan penggunaan ke atas barangan dan perkhidmatan sektor $i$ yang digunakan oleh kumpulan isi rumah $h$ untuk sebelum dan selepas pelaksanaan cukai baharu, $\Delta\hat{P}$ merujuk kepada matrik diagonal perubahan harga *normalized* sektor *i* yang diperolehi daripada model (7) dan mewakili matrik perbelanjaan penggunaan ke atas sektor *i* oleh kumpulan isi rumah *h*. Data perbelanjaan penggunaan isi rumah terdapat di dalam jadual input-output tetapi tiada pecahan perbelanjaan mengikut kumpulan isi rumah tertentu boleh diperolehi. Untuk memperinci impak ke atas perbelanjaan isi rumah ini kepada kumpulan etnik dan kelas pendapatan, data Tinjauan Perbelanjaan Isi rumah (HES) bagi tahun 2010 digunakan.

Model (8) mengandaikan struktur perbelanjaan isi rumah tidak berubah selepas pelaksanaan CBP. Ini bermakna kesan penggantian akibat daripada peningkatan harga selepas CBP tidak diambilkira. Andaian ini adalah realistik dalam jangkamasa panjang kerana struktur perbelanjaan isi rumah dalam jangka masa lima tahun stabil. Bagi tujuan verifikasi, kami telah membandingkan struktur perbelanjaan isi rumah bagi tahun 2005 dan 2010 dengan menggunakan data HES (data 2005 dikira dengan menggunakan harga malar 2010). Dapatan menunjukkan struktur perbelanjaan isi rumah hanyalah berubah di antara 1.96% (tertinggi) dan 0.21% (terendah).

## DATA

Terdapat empat jenis data utama yang digunakan oleh kajian ini. Pertama, jadual input-output Malaysia tahun 2010 yang diterbitkan oleh Jabatan Perangkaan Malaysia. Dengan menggunakan jadual input-output terkini yang diterbitkan pada penghujung tahun 2014, 124 sektor ekonomi yang terdapat di Malaysia telah diliputi. Kedua, data Tinjauan Perbelanjaan Isi Rumah (HES) 2010 juga diperlukan untuk menilai impak CBP ke atas kos sara hidup isi rumah. Secara ringkasnya, HES merupakan tinjauan yang dilakukan ke atas isi rumah di seluruh Malaysia untuk merekod perbelanjaan mereka. Secara keseluruhannya, data HES yang diberikan oleh Jabatan Perangkaan Malaysia terdiri daripada 21,641 isi rumah. Seterusnya, data yang ketiga yang diperlukan adalah pecahan output mengikut aktiviti ekonomi yang boleh didapati melalui laporan banci ekonomi. Data ini digunakan untuk analisis impak kos pengeluaran. Untuk melengkapkan data yang diperlukan, senarai kadar cukai CBP mengikut barangan dan perkhidmatan pula diperoleh daripada pihak Jabatan Kastam Diraja Malaysia (2015). Senarai ini diperlukan untuk menentukan jenis barangan dan perkhidmatan yang tertakluk di bawah kategori pembekalan berkadar sifar dan pembekalan dikecualikan.

Pemadanan dan harmonisasi data perlu dilakukan kerana jadual input-output, HES dan senarai barangan dan perkhidmatan yang diperolehi daripada Jabatan Kastam Diraja Malaysia menggunakan klasifikasi yang berbeza. Pertama, pemadanan dilakukan di antara senarai barangan dan perkhidmatan yang dikategorikan dibawah pembekalan berkadar sifar dan pembekalan dikecualikan dengan senarai barangan dan perkhidmatan HES yang menggunakan *Classification of Individual Consumption According to Purpose* (COICOP). Pemadanan ini bertujuan untuk mengkelaskan barangan dan perkhidmatan berdasarkan aktiviti kepenggunaan isi rumah. Seterusnya, senarai COICOP tersebut telah dipadankan dengan senarai sektor di dalam jadual input-output yang menggunakan klasifikasi MSIC.

## DAPATAN KAJIAN DAN PERBINCANGAN

Perbincangan dapatan kajian dibahagi kepada dua bahagian. Bahagian 4.1 memperinci impak pelaksanaan CBP ke atas kos pengeluaran. Bahagian 4.2 pula melanjutkan lagi analisis pada Bahagian 4.1 dengan membuat penilaian sejauhmana perubahan kos pengeluaran memberi impak terhadap perbelanjaan isi rumah.

## IMPAK KOS PENGELUARAN

Di bawah sistem percukaian CBP, cukai dikenakan kepada pengeluar di sepanjang rantaian bekalan dan pengeluar pula akan melepaskan cukai tersebut kepada pengguna dengan membuat perubahan kepada harga akhir. Konsep ini konsisten dengan tatacara *price pass-through* di dalam model-harga input-output, di mana perubahan kos pengeluaran akan diterjemahkan kepada perubahan harga barangan dan perkhidmatan. Sebelum membincangkan dapatan kajian, perlu ditegaskan bahawa analisis impak CBP ke atas kos pengeluaran dilaksanakan dengan mengandaikan kesemua pertubuhan perniagaan mematuhi sepenuhnya sistem percukaian CBP. Ini bermakna bahawa sistem CJP diganti sepenuhnya dengan sistem CBP di sepanjang rantaian bekalan. Kajian ini mengambil maklum bahawa kadar pematuhan kepada sistem percukaian CBP selalunya adalah rendah pada peringkat awal pelaksanaanya dan akan meningkat dalam jangka masa panjang. Permodelan ke atas kadar pematuhan sistem percukaian CBP yang berbeza-beza tidak dapat dilaksanakan kerana kekangan data. Secara spesifiknya, kami memerlukan data perolehan (output) dan perbelanjaan (input) yang terperinci bagi setiap pertubuhan yang berdaftar di bawah sistem CBP.

Berpandukan kepada maklumat yang disalurkan oleh Jabatan Kastam Diraja Malaysia (2015), terdapat sebanyak 23 sektor di bawah klasifikasi input-output adalah terdiri daripada pembekalan berkadar sifar dan 10 sektor di bawah pembekalan dikecualikan. Perlu diingatkan di sini bahawa setiap sektor di dalam jadual input-output adalah terdiri daripada sekumpulan aktiviti-aktiviti ekonomi yang dikelaskan mengikut MSIC. Oleh yang demikian, setiap sektor ekonomi tidak semestinya tergolong di bawah pembekalan berkadar standard (6%), pembekalan berkadar sifar atau pembekalan dikecualikan sahaja kerana berkemungkinan terdapat satu atau lebih aktiviti ekonomi di dalam sektor tersebut yang tergolong

JADUAL 2. Sektor yang mengalami peningkatan tingkat harga (%)

| Sektor | (%) | Sektor | (%) |
|---|---|---|---|
| Padi | 4.7 | Elektrik dan Gas | 2.4 |
| Penyelidikan dan Pembangunan | 4.5 | Penapisan Petroleum | 2.0 |
| Pemilikan Kediaman | 4.4 | Institusi Swasta bukan Berorientasi Keuntungan | 1.7 |
| Buah-buahan | 4.2 | Pentadbiran Awam | 1.7 |
| Tanaman Bunga | 4.1 | Pertanian Lain | 1.5 |
| Pentadbiran Awam Lain | 4.1 | Pembaikan dan Penyelenggaraan | 1.2 |
| Perkhidmatan Operasi Lebuhraya, Jambatan dan Terowong | 4.0 | Profesional | 1.2 |
| Tanaman makanan | 4.0 | Kediaman | 1.0 |
| Kerja Air | 3.9 | Pembentungan, Pengurusan Sisa dan Aktiviti Pemulihan | 0.9 |
| Minyak Mentah dan Gas Asli | 3.7 | Bukan Kediaman | 0.8 |
| Perkhidmatan Swasta Lain | 3.6 | Bank | 0.7 |
| Penggalian Batu, Tanah Liat dan Pasir | 3.6 | Kejuruteraan Awam | 0.6 |
| Pendidikan | 3.5 | Kerja Pertukangan Khas | 0.6 |
| Sayur-sayuran | 3.4 | Penerima & Pemancar Televisyen dan Radio, Rakaman Bunyi atau Video atau Perkakasan Rakaman Semula dan Barang-barang Berkaitan | 0.5 |
| Perlombongan Bijih Timah | 3.1 | Aktiviti Penerbitan | 0.3 |
| Pengilangan Bijirin | 2.6 | Jentera Kegunaan Khusus | 0.2 |
| Perlombongan dan Penggalian Lain | 2.4 | Tanah Liat dan Seramik | 0.2 |
|  |  | **Purata keseluruhan (34 sektor)** | **2.0** |

di bawah mana-mana kadar percukaian. Sebagai contoh, terdapat 26 aktiviti ekonomi yang diklasifikasi di bawah sektor perikanan (sektor nombor 12 di dalam jadual input-output) tetapi hanya empat aktiviti sahaja yang dikenakan cukai pembekalan berkadar sifar dan selebihnya dikenakan pembekalan berkadar standard. Dengan menggunakan maklumat ini, model harga input-output telah dilaksanakan dengan pengubahsuaian (rujuk Lampiran 1).

Impak pelaksanaan cukai CBP ke atas harga 124 sektor diberikan pada Jadual 2 bagi barangan dan perkhidmatan yang mengalami peningkatan harga dan Jadual 3 bagi barangan dan perkhidmatan yang mengalami penurunan harga. Daripada 124 jenis barangan dan perkhidmatan, 73% dijangka mengalami penurunan tingkat harga dengan kadar purata penurunan sebanyak 7.4%. Selebihnya iaitu 27% pula telah menunjukkan peningkatan harga dengan kadar purata peningkatan sebanyak 2.0%. Secara relatifnya, tingkat harga barangan dan perkhidmatan di dalam ekonomi akan menurun sebanyak 5.4% selepas pelaksanaan CBP.

Sistem percukaian CBP berpotensi menurunkan tingkat harga kerana cukai CJP yang dilaksanakan sebelum ini mempunyai kadar percukaian yang lebih tinggi berbanding CBP. Perlu ditegaskan di sini bahawa di bawah sistem CBP, cukai yang dikenakan adalah berdasarkan nilai ditambah manakala cukai CJP dikenakan berdasarkan output. Rajah 2 di bawah menunjukkan perbezaan di antara pekali cukai CBP kepada output dan pekali cukai CJP kepada output. Pekali-pekali tersebut diperolehi dengan mengira nisbah cukai kepada output. Rajah 2 jelas sekali menunjukkan bahawa pekali cukai CBP yang mengikut nilai ditambah lebih kecil berbanding pekali CJP. Hal ini menyumbang kepada penurunan harga selepas pelaksanaan CBP.

JADUAL 3. Sektor yang mengalami peningkatan tingkat harga (%)

| Sektor | (%) | Sektor | (%) |
|---|---|---|---|
| Pengawetan Makanan Laut | 39.9 | Pengangkutan Air | 7.5 |
| Minyak dan Lemak | 32.8 | Kenderaan Bermotor | 6.8 |
| Pengeluaran Tenusu | 25.8 | Keluaran Kayu Lain | 6.7 |
| Restoran | 24.8 | Benang & Pakaian | 5.8 |
| Aktiviti Wayang Gambar, Video dan Televisyen | 18.6 | Wain & Minuman Keras | 5.7 |
| Getah | 17.0 | Baja | 5.7 |

*Penganggaran Impak Pelaksanaan CBP ke atas Kos Pengeluaran dan Kos Sara Hidup di Malaysia* 23JADUAL 3. (*Sambungan*)

| Sektor | (%) | Sektor | (%) |
|---|---|---|---|
| Pengawetan Buah dan Sayur-sayuran | 13.3 | Kepingan Venir, Papan Lapis, Papan Berlamina, Papan Zarah dan Papan dan Papan Panel Lain | 5.6 |
| Manisan | 12.7 | Daging dan Pengeluaran Daging | |
| Percetakan | 11.5 | Institusi Kewangan Lain | 5.5 |
| Keluaran Kilang Roti | 11.3 | Motosikal | 5.3 |
| Minuman Ringan | 11.2 | Pengilangan dan Pengetaman Kayu | 5.0 |
| Bahan Kimia Asas | 10.4 | Tempat Penginapan | 5.0 |
| Institusi Kewangan | 9.9 | Sabun & Bahan Pencuci, Pewangi, Pencuci & Produk Dandanan Diri | 5.0 |
| Kertas dan Keluaran Kertas dan Perabot | 9.8 | Bekas daripada Kayu dan Rotan | 4.5 |
| Makanan Binatang | 9.6 | Pertukangan dan Kerja Kayu Halus | 4.3 |
| ICT & Perkhidmatan Komputer | 9.1 | Tayar | 4.0 |
| Produk Kimia Lain | 8.5 | Keluaran Tembakau | 3.7 |
| Penyiapan Tekstil | 8.4 | Perdagangan Borong & Runcit dan Kenderaan Bermotor | 3.3 |
| Telekomunikasi | 8.0 | Sarung Tangan Getah | 3.2 |
| Perikanan | 7.9 | Perkhidmatan Hiburan dan Rekreasi | 2.9 |
| Prosesan Makanan Lain | 7.8 | Kelengkapan Pengangkutan Lain | 2.9 |
| Perkhidmatan Operasi Pelabuhan dan Lapangan Terbang | 7.6 | Cat & Varnis | 2.9 |
| Prosesan Getah | 2.9 | Mesin Elektrik Lain | 1.4 |
| Keluaran Getah | 2.8 | Jam Tangan dan Jam | 1.2 |
| Perkhidmatan Pengangkutan Lain | 2.7 | Instrumen Optik & Kelengkapan Fotografi | 1.2 |
| Pembinaan Kapal dan Perahu dan Pembuatan Basikal dan 'Invalid Carriages' | 2.6 | Lampu Elektrik dan Kelengkapan Pencahayaan | 1.1 |
| Mesin Elektrik dan perkakas | 2.6 | Pengangkutan Darat | 1.1 |
| Wayar dan Kabel Berpenebat | 2.6 | Pembuatan Lain | 1.1 |
| Keluaran Plastik | 2.4 | Ketenteraman Awam | 1.1 |
| Hartanah | 2.3 | Pakaian | 1.1 |
| Insurans | 2.3 | Pengangkutan Udara | 1.1 |
| Tekstil Lain | 2.2 | Kaca dan Keluaran Kaca | 1.0 |
| Kelapa Sawit | 2.2 | Keluaran Logam Lain Yang Direka | 0.9 |
| Penerbitan | 2.2 | Mesin Pejabat Perakaunan dan Pengira | 0.9 |
| Menyukat, Memeriksa, & Peralatan Pemprosesan Industri | 2.2 | Peranti Semi-Konduktor Injap Elektronik, Tiub & Papan Litar Pencetak | 0.9 |
| Perhubungan | 2.2 | Kasut | 0.7 |
| Perhutanan dan Pembalakan | 2.1 | Menternak Ayam Itik | 0.7 |
| Industri Kulit | 2.1 | Simen, Kapur & Plaster | 0.6 |
| Logam Asas Berharga & Logam Asas Bukan Ferum | 2.0 | Farmaseutikal, Kimia Perubatan dan Produk Botani | 0.5 |
| Kelengkapan Perubatan & Pembedahan & Peralatan Ortopedik | 1.9 | Ternakan Lain | 0.5 |
| Keluaran Besi dan Keluli | 1.7 | Peralatan Domestik | 0.4 |
| Penuangan Logam | 1.7 | Sewa & Pajak | 0.4 |
| Keluaran Struktur Logam | 1.6 | Perkhidmatan Perniagaan | 0.3 |
| Jentera Industri | 1.6 | Jentera Kegunaan Am | 0.1 |
| Konkrit & Produk Mineral Bukan Logam Lain | 1.6 | Kesihatan | 0.0 |
| | | **Purata keseluruhan (90 sektor)** | **7.4** |

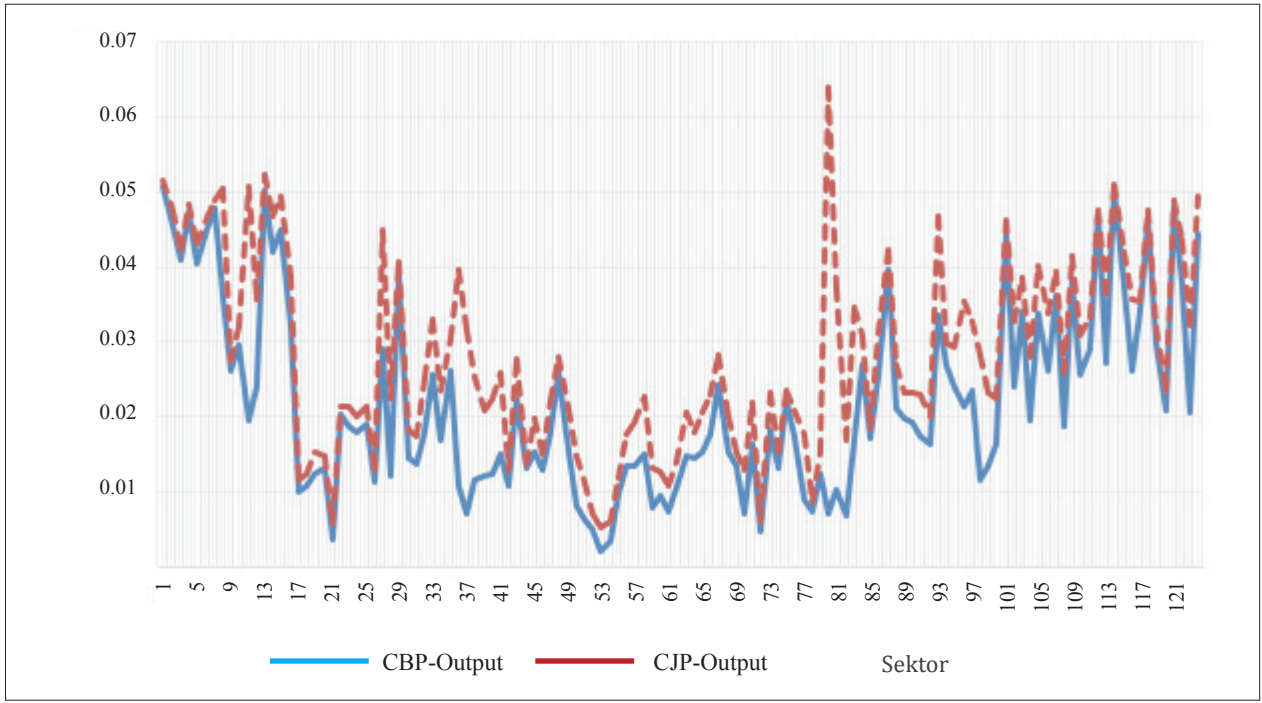

RAJAH 2. Pekali CBP-output dan pekali CJP-output

Perubahan tingkat harga selepas pelaksanaan CBP bagi setiap sektor ditentukan oleh kesalingbergantungan di antara sektor tersebut dengan sektor-sektor lain di dalam ekonomi. Contohnya, harga sektor padi dijangka akan meningkat sebanyak 4.7%, iaitu tertinggi di kalangan sektor yang mengalami peningkatan harga. Sebahagian besar daripada bahan mentah yang digunakan oleh sektor padi adalah dibekalkan oleh sektor pengeluaran baja; perdagangan borong, runcit dan kenderaan bermotor; dan bahan kimia asas. Walau bagaimanapun, perbelanjaan bahan mentah daripada ketiga-tiga sektor ini tidak menyumbang kepada kenaikan harga padi kerana harga ketiga-tiga keluaran tersebut mengalami penurunan. Sebaliknya, kenaikan harga padi disumbang oleh perbelanjaan bahan mentah ke atas sektor-sektor yang mengalami kenaikan harga, terutamanya perkhidmatan profesional; penapisan petroleum; dan elektrik dan gas.

Dalam kalangan sektor-sektor yang mengalami penurunan harga, sektor pengawetan makanan laut dijangka mengalami penurunan harga yang tertinggi, iaitu 40%. Sektor ini membelanja sebahagian besar (iaitu 25%) daripada bahan mentah yang digunakan daripada sektor perikanan. Sektor perikanan pula bukan sahaja mengalami penurunan harga sebanyak 7.9% malah sebahagian besar daripada output sektor ini tergolong di bawah kategori berkadar sifar. Justeru itu, kebergantungan yang kuat di antara sektor pengawetan makanan laut dengan sektor perikanan menyumbang kepada penurunan harga sektor pengawetan makanan laut.

## IMPAK KOS SARA HIDUP ISI RUMAH

Perbincangan pada Bahagian 4.1 telah memperinci impak pelaksanaan percukaian CBP ke atas harga bagi 124 jenis barangan dan perkhidmatan. Bahagian ini pula membincangkan sejauhmana perubahan harga tersebut memberi kesan terhadap perbelanjaan barangan dan perkhidmatan oleh isi rumah di Malaysia. Bagi tujuan pelaporan, kumpulan perbelanjaan ke atas barangan dan perkhidmatan dikelaskan mengikut 12 kategori perbelanjaan, iaitu (i) makanan dan minuman bukan alkohol, (ii) minuman alkohol dan tembakau, (iii) pakaian dan kasut, (iv) perumahan, air, elektrik, gas dan bahan api, (v) hiasan, perkakasan dan penyelenggaraan isi rumah, (vi) kesihatan, (vii) pengangkutan, (viii) komunikasi, (ix) perkhidmatan rekreasi dan kebudayaan, (x) pendidikan, (xi) restoran dan hotel, dan (xii) pelbagai barangan dan perkhidmatan.

Kesan ke atas perbelanjaan isi rumah diukur dengan melihat kepada perbezaan di dalam penggunaan sebelum dan selepas CBP. Jadual 4 memperinci kesan ke atas isi rumah mengikut enam kumpulan kelas pendapatan. Bagi setiap kumpulan pendapatan, lajur pertama merujuk kepada komposisi perbelanjaan isi rumah manakala lajur kedua memberi peratus perubahan perbelanjaan selepas pelaksanaan CBP. Contohnya, isi rumah berpendapatan bawah RM1000 membelanja 33.8% (lajur pertama) daripada jumlah perbelanjaan bulanan (RM692) ke atas kategori i (makanan dan minuman bukan alkohol). Selepas pelaksanaan CBP (lajur kedua), kos perbelanjaan ke atas barangan kategori i telah berkurangan sebanyak

JADUAL 4. Peratus perbelanjaan isi rumah sebelum dan selepas pelaksanaan CBP mengikut kelas pendapatan (%)

| Kumpulan perbelanjaan | < RM1000 | | RM1000-RM1999 | | RM2000-RM2999 | | RM3000-RM3999 | | RM4000-RM4999 | | > RM5000 | |
|---|---|---|---|---|---|---|---|---|---|---|---|---|
| | E | ΔE | E | ΔE | E | ΔE | E | ΔE | E | ΔE | E | ΔE |
| i. Makanan dan minuman bukan alkohol | 33.84 | -1.59 | 27.57 | -7.5 | 22.8 | -7.14 | 19.61 | -5.25 | 18.08 | -3.91 | 13.52 | -10.91 |
| ii. Minuman alkohol dan tembakau | 2.02 | 0.03 | 2.42 | 0.2 | 2.61 | 0.23 | 2.78 | 0.19 | 2.52 | 0.14 | 1.64 | 0.25 |
| iii. Pakaian dan kasut | 3.44 | 0.16 | 3.93 | 0.91 | 3.78 | 0.95 | 3.46 | 0.73 | 3.46 | 0.58 | 3.39 | 2.01 |
| iv. Perumahan, air, elektrik, gas dan bahan api lain | 30.33 | 2.26 | 26.74 | 9.69 | 24.56 | 9.73 | 23.27 | 7.69 | 23.51 | 6.27 | 23.03 | 21.64 |
| v. Hiasan, perkakasan dan penyelengaraan isi rumah | 2.77 | 0.03 | 3.38 | 0.21 | 3.65 | 0.36 | 3.85 | 0.46 | 4.26 | 0.5 | 4.88 | 2.44 |
| vi. Kesihatan | 1.35 | 0.08 | 1.31 | 0.37 | 1.15 | 0.36 | 1.4 | 0.36 | 1.21 | 0.25 | 1.56 | 1.14 |
| vii. Pengangkutan | 6.41 | 0.35 | 11.73 | 3.12 | 15.05 | 4.38 | 16.42 | 3.77 | 17.1 | 3.23 | 17.32 | 9.93 |
| viii. Komunikasi | 3.19 | -0.04 | 4.07 | -0.3 | 4.98 | -0.41 | 5.66 | -0.38 | 6.16 | -0.33 | 6.33 | -1.22 |
| ix. Perkhidmatan rekreasi dan kebudayaan | 1.97 | 0.05 | 3.08 | 0.51 | 4.02 | 0.77 | 4.63 | 0.81 | 4.76 | 0.65 | 6 | 3.27 |
| x. Pendidikan | 0.73 | 0.07 | 0.92 | 0.41 | 1.35 | 0.66 | 1.3 | 0.53 | 1.36 | 0.45 | 2.05 | 2.38 |
| xi. Restoran dan hotel | 8.61 | -1.72 | 7.37 | -9.73 | 8.07 | -11.6 | 8.9 | -10.62 | 8.87 | -8.36 | 9.54 | -31.03 |
| xiii. Pelbagai barang dan perkhidmatan | 5.35 | 0.31 | 7.47 | 2.1 | 7.99 | 2.36 | 8.71 | 2.13 | 8.71 | 1.67 | 10.73 | 7.25 |
| Perbelanjaan purata bulanan | RM692 (-5.45) | | RM1,479 (-5.46) | | RM2,437 (-5.35) | | RM3,433 (-5.38) | | RM4,449 (-5.17) | | RM7,517 (-4.95) | |

*Nota*: E = peratus perbelanjaan isi rumah sebelum CBP, ΔE = peratus perubahan perbelanjaan isi rumah selepas CBP. Angka di dalam kurungan menunjukkan peratus perubahan perbelanjaan purata (nilai negatif menunjukkan peningkatan kuasa beli isi rumah dan sebaliknya).

1.6% (ini juga bermakna telah berlaku peningkatan kuasa beli sebanyak 1.6%) akibat penurunan harga barangan ini.

Secara keseluruhannya, dapatan kajian menjangkakan akan berlaku penurunan dalam kos perbelanjaan ke atas barangan dan perkhidmatan kategori i (makanan dan minuman bukan alkohol), viii (komunikasi) dan xi (restoran dan hotel). Penurunan harga bagi ketiga-tiga jenis barangan dan perkhidmatan ini dinikmati oleh keenam-enam kelas pendapatan. Perbandingan corak perbelanjaan di antara golongan berpendapatan rendah dan tinggi mendapati sistem percukaian CBP berpotensi membawa manfaat kepada golongan yang berpendapatan rendah. Kos perbelanjaan bagi isi rumah berpendapatan bawah RM1000, RM1000-RM1999 dan RM2000-RM2999 berkurangan di antara 5.4% sehingga 5.5% manakala kos perbelanjaan mereka yang berpendapatan RM3000-RM3999, RM4000-RM4999 dan atas RM5000 dijangka menurun di antara 5.0% dan 5.4%. Perbelanjaan kategori manakah yang menyumbang kepada perbezaan ini?

Bagi kumpulan isi rumah yang berpendapatan bawah RM3000, mereka memperuntukkan perbelanjaan yang lebih besar bagi barangan dan perkhidmatan kategori i (makanan dan minuman bukan alkohol). Perkara ini jelas membuktikan bahawa corak perbelanjaan untuk golongan yang berpendapatan bawah daripada RM3000 lebih tertumpu kepada pembelian keperluan asas. Dengan pelaksanaan CBP, kos yang perlu ditanggung untuk membeli barangan dan perkhidmatan kategori i akan menjadi lebih rendah dengan kadar jangkaan peratusan penurunan sebanyak 1.59%, 7.50% dan 7.14% bagi ketiga-tiga kumpulan kelas pendapatan tersebut. Berbanding dengan kumpulan isi rumah berpendapatan bawah RM3000, kumpulan isi rumah berpendapatan melebihi RM3000 memperuntukan sebahagian besar perbelanjaan mereka ke atas barangan dan perkhidmatan kategori iv (perumahan, air, elektrik, gas dan bahan api). Peningkatan dalam kos bagi perbelanjaan untuk barangan dan perkhidmatan kategori ini dijangka akan berlaku dengan peratusan peningkatan masing-masing sebanyak 7.7%, 6.3% dan 21.6%.

Jadual 5 memperinci kesan perubahan perbelanjaan isi rumah mengikut kumpulan etnik. Perbelanjaan purata bulanan isi rumah etnik Bumiputera adalah sebanyak RM2046, lebih rendah berbanding dengan perbelanjaan purata bulanan etnik Cina dan India yang masing-masing berjumlah RM2775 dan RM2191. Selepas pelaksanaan CBP, kos perbelanjaan isi rumah etnik Bumiputera telah berkurangan lebih tinggi berbanding dengan etnik Cina dan India, iaitu masing-masing 5.3%, 5.1% dan 5.0%. Persoalan di sini ialah perubahan kategori perbelanjaan manakah yang menyumbang kepada peningkatan kuasa beli etnik Bumiputera.

Jika diteliti pada Jadual 5, kesemua kumpulan etnik memperuntukkan sebahagian besar perbelanjaan mereka ke atas empat kumpulan berikut: kategori i (makanan dan minuman bukan alkohol), kategori iv (perumahan, air, elektrik, gas dan bahan api), kategori vii (pengangkutan) dan kategori xi (restoran dan hotel). Kos perbelanjaan selepas pelaksanaan CBP untuk kategori vii meningkat manakala kategori xi menurun. Tetapi perubahan kos perbelanjaan kedua-dua kategori ini tidak menunjukkan perubahan ketara di antara kumpulan etnik. Sebaliknya,

JADUAL 5. Peratus perbelanjaan isi rumah sebelum dan selepas pelaksanaan CBP mengikut kumpulan etnik (%)

| Kumpulan perbelanjaan | Bumiputera | | Cina | | India | | Lain-lain | |
|---|---|---|---|---|---|---|---|---|
| | E | ΔE | E | ΔE | E | ΔE | E | ΔE |
| i. Makanan dan minuman bukan alkohol | 22.02 | -1.20 | 16.84 | -0.97 | 19.25 | -1.24 | 25.03 | -1.14 |
| ii. Minuman alkohol dan tembakau | 1.92 | 0.03 | 2.37 | 0.01 | 2.17 | 0.02 | 3.87 | 0.08 |
| iii. Pakaian dan kasut | 3.94 | 0.17 | 2.64 | 0.11 | 2.95 | 0.12 | 3.24 | 0.16 |
| iv. Perumahan, air, elektrik, gas dan bahan api lain | 20.82 | 1.52 | 24.89 | 1.77 | 24.29 | 1.70 | 25.65 | 2.00 |
| v. Hiasan, perkakasan dan penyelengaraan isi rumah | 4.44 | 0.11 | 3.65 | 0.11 | 3.65 | 0.08 | 2.98 | 0.10 |
| vi. Kesihatan | 1.19 | 0.07 | 1.58 | 0.08 | 1.50 | 0.08 | 0.71 | 0.04 |
| vii. Pengangkutan | 16.47 | 0.72 | 13.58 | 0.69 | 13.30 | 0.67 | 7.81 | 0.43 |
| viii. Komunikasi | 5.26 | -0.09 | 6.22 | -0.11 | 5.66 | -0.11 | 6.40 | -0.07 |
| ix. Perkhidmatan rekreasi dan kebudayaan | 3.83 | 0.16 | 5.85 | 0.17 | 5.59 | 0.17 | 3.65 | 0.19 |
| x. Pendidikan | 1.20 | 0.11 | 1.80 | 0.16 | 1.84 | 0.16 | 0.61 | 0.06 |
| xi. Restoran dan hotel | 10.28 | -2.05 | 12.00 | -2.44 | 10.47 | -2.13 | 11.77 | -2.35 |
| xii. Pelbagai barang dan perkhidmatan | 8.64 | 0.46 | 8.59 | 0.41 | 9.34 | 0.47 | 8.28 | 0.50 |
| Perbelanjaan purata bulanan | RM 2,046 (-5.25) | | RM 2,775 (-5.06) | | RM 2,191 (-4.99) | | RM 1,831 (-5.71) | |

*Nota*: E = peratus perbelanjaan isi rumah sebelum CBP, ΔE = peratus perubahan perbelanjaan isi rumah selepas CBP. Angka di dalam kurungan menunjukkan peratus perubahan perbelanjaan purata (nilai negatif menunjukkan peningkatan kuasa beli isi rumah dan sebaliknya).

JADUAL 6. Jurang penggunaan sebelum dan selepas pelaksanaan CBP

| | Asas (RM) | Pasca-CBP (RM) | Perubahan (%) | Jurang-penggunaan | |
|---|---|---|---|---|---|
| | | | | Asas | Pasca-CBP |
| | (1) | (2) | (3) | (4) | (5) |
| *Kumpulan etnik* | | | | | |
| Bumiputera | 2,046 | 2,153 | 5.25 | 1.000 | 1.000 |
| Cina | 2,775 | 2,915 | 5.06 | 1.356 | 1.354 |
| India | 2,191 | 2,300 | 4.99 | 1.071 | 1.068 |
| Lain-lain | 1,831 | 1,936 | 5.71 | 0.895 | 0.899 |
| *Kumpulan pendapatan* | | | | | |
| <RM1,000 (bawah RM1,000) | 692 | 730 | 5.45 | 1.000 | 1.000 |
| RM1,000 – RM1,999 | 1,479 | 1,560 | 5.46 | 2.137 | 2.137 |
| RM2,000 – RM2,999 | 2,437 | 2,567 | 5.35 | 3.522 | 3.518 |
| RM3,000 – RM3,999 | 3,433 | 3,618 | 5.38 | 4.961 | 4.958 |
| RM4,000 – RM4,999 | 4,449 | 4,679 | 5.17 | 6.429 | 6.412 |
| >RM5,000 (melebihi RM5,000) | 7,517 | 7,889 | 4.95 | 10.863 | 10.811 |

perubahan ketara boleh dilihat pada kategori i dan ii, dan didapati menyumbang kepada perbezaan pengurangan kos perbelanjaan di antara kumpulan etnik.

Perbelanjaan etnik Bumiputera ke atas kategori i, 5.2% dan 2.8% lebih tinggi berbanding etnik Cina dan India. Selepas pelaksanaan CBP, didapati terdapat pengurangan kos perbelanjaan sebanyak 1.20%, 0.97% dan 1.24% bagi etnik Bumiputera, Cina dan India bagi kategori ini. Secara relatifnya, etnik Bumiputera mendapat manfaat lebih besar daripada pengurangan kos perbelanjaan kategori i. Kos perbelanjaan kategori iv menunjukkan peningkatan tetapi kadar peningkatan lebih besar dialami oleh etnik Cina dan India (masing-masing 1.77% dan 1.70%) berbanding Bumiputera (1.52%). Oleh kerana etnik Cina dan India memperuntukan perbelanjaan lebih tinggi ke atas kategori ini (iaitu 4.07% dan 3.47% lebih tinggi daripada Bumiputera), maka peningkatan kos perbelanjaan yang lebih ketara dialami oleh mereka.

Keseluruhannya, analisis mendapati pelaksanaan CBP berpotensi mengurangkan kos perbelanjaan isi rumah yang berpendapatan rendah (iaitu kumpulan berpendapatan di bawah RM3000) dan etnik Bumiputera lebih tinggi berbanding isi rumah berpendapatan tinggi. Jadi, wujud kecenderungan CBP untuk mengecilkan jurang penggunaan isi rumah. Di dalam Jadual 6, lajur (1) merujuk kepada purata perbelanjaan bulanan sebelum pelaksanaan CBP manakala lajur (2) mewakili purata perbelanjaan bulanan selepas pelaksanaan CBP. Lajur (3) memberi kadar peratusan perubahan perbelanjaan bulanan selepas pelaksanaan CBP. Contohnya, peratusan perubahan perbelanjaan untuk etnik Bumiputera ialah 5.25% bermaksud kuasa beli etnik Bumiputera ke atas perbelanjaan semasa telah meningkat sebanyak 5.25% selepas CBP akibat daripada penurunan tingkat harga barangan dan perkhidmatan. Lajur (4) mewakili kadar jurang penggunaan dengan mengambil perbelanjaan purata Bumiputera sebagai asas perbandingan di antara kumpulan etnik. Jurang pendapatan kelas pendapatan isi rumah mengambil perbelanjaan purata kumpulan <RM1000 sebagai asas perbandingan. Contohnya, kadar jurang penggunaan etnik Cina sebelum CBP ialah 1.356 bermaksud setiap Ringgit perbelanjaan oleh etnik Bumiputera bersamaan dengan 1.356 Ringgit perbelanjaan etnik Cina (jadi kuasa beli etnik Cina 36% melebihi kuasa beli Bumiputera). Jurang penggunaan untuk kelas pendapatan boleh ditakrifkan seperti jurang penggunaan etnik.

Dapatan analisis pada Jadual 6 menunjukkan keberkesanan CBP sebagai alat kepada dasar pengagihan. Didapati pelaksanaan CBP berpotensi untuk mengurangkan jurang penggunaan Bumiputera-Cina sebanyak 0.15% daripada 1.356 kepada 1.354 dan sebanyak 0.28% daripada 1.071 kepada 1.068 untuk jurang penggunaan Bumiputera-India. Jurang penggunaan di antara kelas pendapatan juga menunjukkan kesan positif (pengurangan jurang). Contohnya, jurang penggunaan di antara kumpulan isi rumah berpendapatan <RM1,000 dan >RM5,000 menguncup sebanyak 0.48% daripada 10.863 kepada 10.811. Walaupun kadar pengurangan di dalam jurang penggunaan adalah kecil, pelaksanaan CBP dilihat mampu menjadi alat kepada dasar pengagihan di Malaysia. Ini konsisten dengan kajian saintifik yang membuktikan bahawa cukai merupakan salah satu alat dasar pengagihan pendapatan yang efektif.

## RUMUSAN DAN IMPLIKASI DASAR

Kertas ini membentangkan hasil kajian impak pelaksanaan sistem percukaian CBP terhadap ekonomi Malaysia. Fokus kajian ialah menganggar impak CBP terhadap kos pengeluaran barangan dan perkhidmatan dan kos sara hidup isi rumah. Model harga input-output telah dibangunkan dengan pengubahsuaian bagi mengambil kira tiga kadar cukai yang berbeza iaitu pembekalan berkadar standard, pembekalan berkadar sifar dan pembekalan dikecualikan. Dapatan kajian utama dapat dirumuskan seperti berikut. Pertama, pelaksanaan CBP berpotensi untuk menurunkan tingkat harga barangan dan perkhidmatan sebanyak 7.4% berbanding hanya 2.0% peningkatan harga yang dijangka. Kedua, hasil daripada penurunan harga sebanyak 5.4% ini telah meningkatkan kuasa beli bagi semua kumpulan isi rumah yang dikaji dengan kumpulan berpendapatan rendah menikmati peningkatan kuasa beli lebih tinggi berbanding dengan kumpulan berpendapatan tinggi. Hasilnya, jurang penggunaan di antara kumpulan berpendapatan rendah dan yang berpendapatan tinggi dapat dikurangkan. Ketiga, perubahan kos perbelanjaan isi rumah di antara kumpulan isi rumah berpendapatan rendah dan yang berpendapatan tinggi diterangkan oleh perubahan kos perbelanjaan ke atas dua jenis barangan berikut: (i) makanan dan minuman bukan alkohol, dan (ii) perumahan, air, elektrik, gas dan bahan api.

Walau bagaimanapun, dapatan kajian ini perlu ditafsir secara berhati-hati dengan mengambil kira dua faktor berikut. Pertama, analisis mengandaikan kesemua sektor di dalam ekonomi mematuhi dan melaksana sistem percukaian CBP tanpa melibatkan sebarang penyelewengan (ini bermakna kesemua pertubuhan perniagaan mematuhi sepenuhnya sistem CBP). Kedua, analisis harga hanya mengambil kira pembolehubah CBP sahaja manakala pengaruh dasar-dasar tekanan harga lain seperti upah minimum, rasionalisasi subsidi petroleum dan penyusutan nilai matawang Ringgit tidak diliputi, *ceteris paribus*. Lebih menyukarkan lagi apabila pengaruh ketiga-tiga dasar di atas sukar untuk dikawal dan dipengaruhi oleh faktor-faktor luaran. Tinjauan literatur jelas sekali menunjukkan upah minimum, rasionalisasi subsidi petroleum dan penyusutan nilai matawang Ringgit cenderung untuk meningkatkan tingkat harga di dalam ekonomi. Umpamanya, Saari, Shuja' dan Abdul Rahman (2013) mendapati bahawa langkah rasionalisasi subsidi bahan api akan membawa kepada peningkatan kos input bahan api yang kemudiannya akan disalurkan kepada pengguna melalui kenaikan harga barangan. Saari, Hassan dan Said (2013) juga mendapati dasar upah minimum berpotensi untuk meningkatkan kadar kos buruh keseluruhan sektor ekonomi sebanyak 6.4% yang seterusnya membawa kepada kenaikan harga secara purata sebanyak 1.8%. Ibrahim dan Aziz (2003) pula mendapati kejatuhan nilai matawang Ringgit turut meningkatkan kos pengeluaran melalui peningkatan harga bagi input (input modal dan input perantaraan) yang diimport daripada luar negara. Justeru itu, kedua-dua faktor di atas dapat memberi penjelasan mengapa dapatan kajian ini berbeza dengan keadaan sebenar yang berlaku di dalam ekonomi masa kini.

Dapatan kajian ini mendapati CBP berpotensi untuk menurunkan tingkat harga jika kesemua pertubuhan perniagaan mematuhi sepenuhnya sistem CBP. Justeru itu, cadangan pelan tindakan perlu terarah kepada usaha untuk mengalakkan penyertaan pertubuhan perniagaan ke dalam sistem CBP. Penyertaan pertubuhan perniagaan ke dalam sistem CBP boleh dipertingkatkan melalui penguatkuasaan. Penguatkuasaan akan menjadi lebih mudah jika pihak berwajib memahami struktur pasaran yang wujud di Malaysia. Sumber-sumber yang dimiliki oleh pihak penguatkuasaan seperti pegawai penguatkuasaan adalah terhad. Justeru itu, pemantauan kepada pertubuhan yang bersifat monopolistik dan oligopolistik dapat membantu kepada kadar penyertaan pertubuhan-pertubuhan perniagaan lain ke dalam sistem CBP. Penguatkuasaan undang-undang lain yang berkaitan seperti Akta Persaingan 2010 juga perlu dipantau untuk mengukuhkan pelaksanaan CBP.

Asan Ali Golam Hassan
International Business School
Universiti Teknologi Malaysia
Jalan Sultan Yahya Petra (Jalan Semarak)
54100 Kuala Lumpur, Malaysia
Email: asanali@ibs.utm.my

Mohd Yusof Saari*
Institut Kajian Dasar Pertanian dan Makanan
Universiti Putra Malaysia
Putra Infoport
43400 Serdang, Selangor, Malaysia
E-mail : mysaari@gmail.com

Chakrin Utit
Institut Kajian Dasar Pertanian dan Makanan
Universiti Putra Malaysia
Putra Infoport
43400 Serdang, Selangor, Malaysia
E-mail : uchakrin@gmail.com

Azman Hassan
Jabatan Ekonomi
Fakulti Ekonomi dan Pengurusan
Universiti Putra Malaysia
43400 Serdang, Selangor, Malaysia
E-mail : azmanhs@upm.edu.my

Mukaramah Harun
Pusat Pengajian Ekonomi, Kewangan dan Perbankan
Universiti Utara Malaysia
06010 Sintok, Kedah, Malaysia
E-mail : mukaramah@uum.edu.my

*Corresponding author

LAMPIRAN 1. MODEL HARGA INPUT-OUTPUT UNTUK PERCUKAIAN CBP DENGAN PELBAGAI KADAR

Contoh pengiraan berikut dilakukan dengan menggunakan jadual input-output yang diaggregat kepada sektor pertanian, industri dan perkhidmatan, dan diberi pekali input-output seperti berikut:

|  | Transposisi | | | Kos eksogenous | | | Total |
| --- | --- | --- | --- | --- | --- | --- | --- |
|  | Agr | Ind | Ser | Tax | Vad | Imp |  |
| Pertanian (Agr) | 0.06 | 0.11 | 0.12 | 0.01 | 0.60 | 0.10 | 1.00 |
| Industri (Ind) | 0.07 | 0.25 | 0.12 | 0.01 | 0.27 | 0.28 | 1.00 |
| Perkhidmatan (Ser) | 0.01 | 0.12 | 0.28 | 0.01 | 0.48 | 0.10 | 1.00 |

*Nota*: Tax = cukai tidak langsung, Vad = nilai ditambah dan Imp = import

Bahagian metodologi telah menerangkan bagaimana permodelan CBP dilakukan di dalam model harga input-output. Ianya dirumuskan seperti berikut:

$$\Delta \mathbf{p} = (\mathbf{I} - \mathbf{A'}\hat{\mathbf{B}})^{-1}$$

Jika diandaikan sektor Pertanian (Agr) adalah berkadar sifar, dan dengan diberi matrik seperti di dalam jadual di atas, maka $\mathbf{A'}\hat{\mathbf{B}}$ boleh dibentuk seperti berikut:

$$\mathbf{A'}\hat{\mathbf{B}} = \begin{bmatrix} 0.06 & 0.11 & 0.12 \\ 0.07 & 0.25 & 0.12 \\ 0.01 & 0.12 & 0.28 \end{bmatrix} \begin{bmatrix} 0 & 0 & 0 \\ 0 & 1 & 0 \\ 0 & 0 & 1 \end{bmatrix} = \begin{bmatrix} 0.00 & 0.11 & 0.12 \\ 0.00 & 0.25 & 0.12 \\ 0.00 & 0.12 & 0.28 \end{bmatrix}$$

Pada matrik **B**, *Off-diagonal* untuk sektor Pertanian (Agr) diletakkan nilai 0 kerana sektor berkenaan adalah pembekalan berkadar sifar manakala 1 untuk sektor Industri (Ind) dan Perkhidmatan (Ser) kerana pembekalan berkadar standard. Jadi matrik $(\mathbf{I} - \mathbf{A'}\hat{\mathbf{B}})^{-1}$ yang telah dikira adalah seperti berikut:

$$(\mathbf{I} - \mathbf{A'}\hat{\mathbf{B}})^{-1} = \begin{bmatrix} 1.00 & 0.18 & 0.20 \\ 0.00 & 1.36 & 0.23 \\ 0.00 & 0.22 & 1.42 \end{bmatrix}$$

Perlu diingatkan bahawa sesebuah sektor di dalam jadual input-output adalah terdiri daripada beberapa komoditi yang dikumpul bersama mengikut klasifikasi MSIC. Wujud kemungkinan bahawa sebahagian komoditi di dalam sektor Pertanian umpamanya terdiri daripada pembekalan berkadar sifar manakala sebahagian lagi pembekalan berkadar standard. Jika kita andaikan 50% daripada nilai output sektor Pertanian (Agr) terdiri daripada pembekalan berkadar sifar manakala 50% nilai output lagi terdiri daripada pembekalan berkadar standard, maka diberi seperti berikut:

$$\mathbf{A'}\hat{\mathbf{B}} = \begin{bmatrix} 0.06 & 0.11 & 0.12 \\ 0.07 & 0.25 & 0.12 \\ 0.01 & 0.12 & 0.28 \end{bmatrix} \begin{bmatrix} 0.5 & 0 & 0 \\ 0 & 1 & 0 \\ 0 & 0 & 1 \end{bmatrix} = \begin{bmatrix} 0.03 & 0.11 & 0.12 \\ 0.04 & 0.25 & 0.12 \\ 0.01 & 0.12 & 0.28 \end{bmatrix}$$

*Off-diagonal* matrik $\hat{\mathbf{B}}$ untuk sektor Pertanian (Agr) diletakkan 0.5 kerana 50% daripada nilai output daripada sektor ini adalah pembekalan berkadar sifar (jika 30% berkadar sifar maka 0.3). Jadi matrik $(\mathbf{I} - \mathbf{A'}\hat{\mathbf{B}})^{-1}$ yang telah dikira adalah seperti berikut:

$$(\mathbf{I} - \mathbf{A'}\hat{\mathbf{B}})^{-1} = \begin{bmatrix} 1.04 & 0.18 & 0.20 \\ 0.05 & 1.36 & 0.24 \\ 0.02 & 0.23 & 1.43 \end{bmatrix}$$